\documentstyle[12pt]{article}

\newcommand{\eq}{\begin{equation}}
\newcommand{\en}{\end{equation}}
\newcommand{\eqa}{\begin{eqnarray}}
\newcommand{\ena}{\end{eqnarray}}

\begin{document}

\hskip 11.5cm \vbox{\hbox{DFTT 43/95}\hbox{August 1995}}
\vskip 0.4cm
\centerline{\Large\bf Calogero-Sutherland techniques in the physics}
\centerline{\Large\bf of disordered  wires}
\vskip 1cm
\centerline{M. Caselle
\footnote{caselle@to.infn.it}}
\vskip .3cm
\centerline{\sl Istituto Nazionale di Fisica Nucleare, Sezione di Torino}
\centerline{\sl  Dipartimento di Fisica
Teorica dell'Universit\`a di Torino}
\centerline{\sl via P.Giuria 1, I-10125 Turin,Italy}
\vskip .5cm

\vskip 1.5cm

\begin{abstract}
We discuss the connection between the random matrix approach to
disordered wires and the Calogero-Sutherland models. We show that
different choices of random matrix ensembles correspond to different
classes of CS models. In particular, the standard transfer matrix
ensembles correspond to CS model with sinh-type interaction, constructed
according to the $C_n$ root lattice pattern.
  By exploiting
this relation, and by using some known properties of the zonal spherical
functions on symmetric spaces we can obtain several properties of the
 Dorokhov-Mello-Pereyra-Kumar equation, which
describes the evolution of an ensemble of quasi one-dimensional
disordered wires of increasing length $L$. These
 results are in complete agreement with all known properties
of disordered wires. \\
\end{abstract}

\noindent
\vfill
\eject

\newpage

\setcounter{footnote}{0}
\def\thefootnote{\arabic{footnote}}

\section{Introduction}

During last years an increasing interest has been attracted by the
physics of quantum electronic transport in disordered wires~\cite{alw}.
One of the main reasons for this interest lies in the
high degree of universality of some experimental observations.
In particular
it was shown that the variations of the measured conductance as a
function of the magnetic field or of the Fermi energy are independent
of the size and degree of disorder of the sample and have a variance
always of order $e^2/h$. This phenomenon is usually known as
 universal conductance fluctuations
(UCF). Its universality suggests that disordered wires could be
described by some relatively simple Hamiltonian, independent of the
particular model or disorder realization. This approach was pioneered
by Imry~\cite{im} and developed by Muttalib, Pichard and
Stone~\cite{mps}, who suggested to describe UCF by constructing
a Random Matrix Theory (RMT) of quantum transport, in analogy to the
Wigner-Dyson RMT for nuclear energy levels. However it was soon realized
that, besides the obvious analogies, there are some relevant
differences between the Wigner-Dyson ensembles (WDE) of random matrices
 and
those constructed to describe the physics of disordered wires (which we
shall denote in the following as Transfer Matrix ensembles (TME)).
The most interesting feature of these TME is the presence of a
Fokker-Plank type evolution equation, known as DMPK equation (see below)
which can be considered the equivalent in the context of TME of the
Brownian motion approach to  WDE suggested by Dyson~\cite{dys}
(for a discussion of these analogies see~\cite{fp}).

The aim of this contribution is to show that the relationship
between WDE and TME (and the parallel one between Brownian motion and
 DMPK equation) can be well
understood by exploiting their equivalence with the so called
``Calogero-Sutherland'' (CS)  models~\cite{cs} which are quantum integrable
models describing a set of $N$ particles moving along a line (see
below). In particular it will be shown that the Schr\"odinger equation of
the CS models is equivalent to the DMPK equation and that  WDE's,
TME's and the S-matrix ensembles described in~\cite{fp}
correspond to different realizations of the CS models. We shall show
that the common, underlying, mathematical structure of all these models
 is  the theory of Laplace-Beltrami operators on Symmetric Spaces.
By using some recent results on the eigenfunctions of these operators,
which are known as ``zonal spherical functions'' we shall construct
exact (for
the unitary ensemble) or asymptotic (in the other cases) solutions of
the DMPK equation and describe several physical properties of disordered
wires.

This contribution is organized as follows: after a short
introduction to disorder wires and to their TME description (sect.2)
we shall discuss the DMPK equation (sect.3). In Sect. 4 we shall give a
short introduction to the CS models and show the anticipated equivalence
with the TME. Sect.5 will be devoted to the solution of the DMPK
equation and sect.6 to some concluding remarks.

\section{Disordered wires}

The peculiar feature of  disordered wires  is that in these
systems, at low enough temperature, the phase coherence of the
electron's wavefunctions is kept over distances much larger than the
typical mean free path, thus allowing to study several non-trivial
quantum effects. These phenomenon can be studied only at low
temperatures where the {\sl inelastic} electron-phonon scattering (which
changes the phases of electrons in a random way) becomes negligible  and
resistivity is completely dominated by the scattering against random
impurities, which is {\sl elastic} and changes the phases in a
reproducible, deterministic way. The natural theoretical framework
to describe these systems is the Landauer theory~\cite{lan} which
assumes the electrons in thermal equilibrium with the various chemical
potentials in the leads. The disordered wire, with its impurities, is
then regarded as a scattering center for the electrons originating from
the current leads and the conductance $G$ is proportional to the
transmission coefficients of the scattering problem. Within this
approach Fisher and Lee~\cite{fl} proposed the following expression for
the conductance in a two-probe geometry (namely a finite disordered
section of length $L$ and transverse width $W$, to which current is
supplied by two semi-infinite ordered leads):
\eq
G=G_0~{\rm Tr}(tt^\dagger)\equiv G_0\sum_n T_n,~~~~~~~
G_0=\frac{2e^2}{h}.
\label{1a}
\en
where $t$ is  the $N\times N$
transmission matrix of the conductor and
 $T_1,T_2 \cdots T_N$ are the eigenvalues of the product $tt^\dagger$
and are usually called transmission eigenvalues.
$N$ is the number of scattering channels at the Fermi level. $N$ depends
on the width of the wire and even in the narrowest metal wires
 it is of the order of $N\sim 10^4-10^5$ so that for metal wires
a large $N$ approximation will give in general very good results
 (notice however that semiconductor microstructure with very low values
of $N$ can be constructed and studied).
In the following we shall often refer to the dimensionless conductance
$g$, defined as $g \equiv G/G_0$. The transmission matrix $t$ is a
component
of the $2N\times 2N$ scattering matrix $S$ which relates the incoming
flux to the outgoing flux:
\eq
S\left(\begin{array}{c}I \\  I' \end{array}\right)=
\left(\begin{array}{c}O \\  O' \end{array}\right)~~~,
\label{1b}
\en
\eq
S=\left(\begin{array}{cc}r & t \\t' & r' \end{array}\right)~~,
\label{1c}
\en
where I,O,I',O' are $N$ component vectors which describe the incoming
 and outgoing wave amplitudes on the left and right respectively, and
$r$ is the $N\times N$ reflection matrix. Current conservation implies
\eq
|I|^2+|I'|^2=|O|^2+|O|^2~~~,
\label{1d}
\en
which is equivalent to the requirement of unitarity: $S\in U(2N)$.
However it turns out that for the problem that we are studying a much
better parametrization is given by the transfer matrix $M$ which is
defined as:
\eq
M\left(\begin{array}{c}I \\  O \end{array}\right)=
\left(\begin{array}{c}O' \\  I' \end{array}\right)~~~,
\label{1e}
\en

The relation between the transfer matrix and the transmission
eigenvalues becomes clear if one constructs an auxiliary matrix $Q$
defined in terms of $M$ as follows:
\eq
Q=\frac{1}{4}[M^\dagger M +(M^\dagger M)^{-1} -2]
\label{1f}
\en
 The eigenvalues $\{\lambda_i\}$ of $Q$ are non-negative and can be
related to the transmission eigenvalues $T_i$ by
\eq
\lambda_{i}\equiv (1-T_{i})/T_{i}
\label{1g}
\en
The physics of the disordered wires that we are studying will be
completely described if we can obtain the probability distribution
 $P(\{\lambda_i\})$ for the eigenvalues $\lambda_i$ (and consequently
for the $T_i$'s).

To this end let us first study some general properties of the transfer
matrix which are direct consequences of the physical symmetries of the
problem. First of all, it is easy to see that
the same flux conservation constraint eq.(\ref{1d}) discussed above
implies in this case the conservation of a hyperbolic norm:
\eq
M^{\dagger}\Sigma_z M=\Sigma_z
\label{1e2}
\en
with
\eq
\Sigma_z=\left(\begin{array}{cc}{\bf 1} & {\bf 0} \\{\bf 0}
 & {\bf -1} \end{array}\right)~~,
\label{1e3}
\en
where ${\bf 0}$ and ${\bf 1}$ are the zero and unit $N\times N$
matrices. As a consequence of (\ref{1e2}), $M \in SU(N,N)$. The ensemble
of transfer matrices defined in this way is usually called ``unitary
ensemble'' (ensemble IIa in the notation of ref.~\cite{el}).

If the system is also invariant under time reversal symmetry,
$M$ must satisfy a further constraint:
\eq
M^{*}\Sigma_x M=\Sigma_x
\label{1e4}
\en
with
\eq
\Sigma_x=\left(\begin{array}{cc}{\bf 0} & {\bf 1} \\{\bf 1}
 & {\bf 0} \end{array}\right)~~,
\label{1e5}
\en
It is possible to show that the joint application of (\ref{1e2}) and
(\ref{1e4}) implies $M\in SP(2N,{\bf R})$~\cite{mpk}.  The ensemble
of transfer matrices defined in this way is usually called ``orthogonal
ensemble'' (ensemble I in the notation of ref.~\cite{el}).

{}From an experimental point of view it is very simple to control the time
reversal symmetry which is eliminated by the application of an external
magnetic field.

If the disordered wire contains ``magnetic impurities'', namely if the
spin-orbit interaction in the scattering against impurities becomes
important then the spin-rotation symmetry (which was implicitly
assumed in all the above discussion) is not any more conserved. In this
case the two spin components of the electrons must be treated
separately.
Each one of the input and output vectors $I,I',O,O'$ becomes a collection
of $N$ spinors (one for each channel) and each spinor contains the two
spin degrees of freedom. Hence $M$ is in this case a $4N \times 4N$
complex matrix. If only flux conservation is imposed (time reversal
symmetry broken) then $M\in U(2N,2N)$ and we find again the unitary
ensemble described above with the only change: $N\to 2N$
(this difference was kept explicit in  ref.~\cite{el}, where this case
 was denoted as IIb to distinguish it from the spin-rotation symmetric
unitary ensemble IIa).
If time reversal symmetry is conserved (namely if there are magnetic
impurities, but no external magnetic field) then one must impose
the following constraint on $M$~\cite{mc}
(analogous of that of eq.(\ref{1e4}))

\eq
M^*=K M K^{T}
\label{m4}
\en
with
\eq
K=
\left(\begin{array}{cc}\mbox{\bf 0}& \mbox{\bf A} \\ \mbox{\bf A}&
\mbox{\bf 0}
 \end{array}\right)~~~, \nonumber
\en
where ${\bf A}$ is a $2N\times 2N$ block diagonal matrix
\eq
{\bf  A}=\sigma {\bf 1}~,~~~
\sigma=
\left(\begin{array}{cc} 0 & 1 \\ -1 & 0
 \end{array}\right)~.
\en

This constraint implies that $M\in SO^*(4N)$ and defines the so called
symplectic ensemble (ensemble III in the language of ref~\cite{el}).

The second step in order to construct the probability distribution for
the $\lambda_i$'s comes from the identification of the
$\lambda_i$'s themselves as the relevant physical degrees of freedom of
the system. This identification has some very interesting
 group theoretical consequences. In fact
 the choice of the $\lambda_i$'s as relevant physical parameters
induces the following parametrization for $M$ for the orthogonal and
unitary cases~\cite{mpk}:
\eqa
M&=&\left(\begin{array}{cc} u^{(1)} & \mbox{\bf 0} \\
\mbox{\bf 0} & u^{(2)}  \end{array}\right)
\left(\begin{array}{cc} \sqrt{1+{\bf \Lambda}} &
\sqrt{{\bf \Lambda}} \\
 \sqrt{{\bf \Lambda}} &  \sqrt{1+{\bf \Lambda}}
\end{array}\right) \nonumber \\
&\times&
\left(\begin{array}{cc} u^{(3)} & \mbox{\bf 0} \\
\mbox{\bf 0} & u^{(4)}  \end{array}\right)\equiv U\Gamma V
\label{1h}
\ena
where
${\bf \Lambda}$  is a  $N\times N$ real, diagonal,
matrix with entries the eigenvalues $\lambda_1,\lambda_2,
\cdots \lambda_N$ in both cases.
The $u^{(i)}$, $(i=1,2,3,4)$ are 4 independent $N\times N$  unitary
matrices in the unitary case while in the orthogonal case they are
constrained by the relations:
\eq
u^{(2)}=u^{(1)*}, u^{(4)}=u^{(3)*}
\label{1i}
\en

In the symplectic case we have again the same parametrization, if the
various matrices are written in terms of quaternions. Thus in this case
${\bf \Lambda}$ is a $N\times N$ quaternion real, diagonal matrix and the
$u^{(i)}$'s are $N\times N$ quaternion, unitary matrices which again
obey the constraint (\ref{1i}).

In this parametrization we  recognize a $G/H$ coset
structure, where $G$ is the group to which the transfer matrix belongs
and $H$ is the group to which the $U$ and $V$ matrices belong. The coset
structure is immediately evident if we notice that any transformation
$M\to M'=WMW^{-1}$, with $W\in H$ gives again a transfer matrix which
can be decomposed as $M'=U'\Gamma V'$ with $U'=WU$, $V'=VW^{-1}$ and the
{\sl same} matrix $\Gamma$. So the physically relevant parameters
$\lambda_i$ are left unchanged by such transformation and thus belong to
the coset space $G/H$. These cosets are listed in Tab.1 for the three
ensembles in which we are interested.

\begin{table}[htb]
\small
\begin{center}
\begin{tabular}{|c|c|c|c|}
\hline
 $G$ & $H$  &  T & S-R  \\
\hline
$Sp(2N,\mbox{\bf R})$ & $U(N)$ & $y$  & $y$  \\
$SU(N,N)$ & $SU(N)\otimes SU(N)\otimes U(1)$  & $n$ & $y$  \\
$SO^*(4N)$ & $U(2N)$  &  $y$ & $n$  \\
\hline
\end{tabular}
\end{center}
\caption
        {\label{cn}
    $G/H$ cosets for the orthogonal (first line), unitary (second line)
    and symplectic (last line) ensembles.
    In the first two columns the group $G$ and the
    subgroup $H$. In the last two columns the status of the time
reversal (T) and spin rotation (S-R) symmetries respectively (y=
conserved, n=broken).}
\end{table}

 However this is not the end of the story. Looking
at the three particular realizations of the pair $G,H$
 listed in tab.1 we see that in all the three
cases the cosets are actually symmetric spaces (see tab.2).
What is more important,
we recognize in the parametrization (\ref{1h}) the so called ``spherical
decomposition'' of those symmetric spaces
(see for instance~\cite{helg}). This tells us that the $\lambda_i$'s play
the role of (generalized) {\sl radial coordinates} in $G/H$
and  implies that the
$\lambda_i$'s are also invariant under (generalized) angular
transformations in $G/H$ and that only the radial projection of any
given dynamical operator will influence their dynamics.

At this point the only remaining step is to impose some dynamical
principle so as to obtain an ``equation of motion'' for the probability
distribution of the $\lambda_i$'s.

\section{The DMPK equation.}

This program was completed during the eighties, at least in the case of
quasi one-dimensional wires,  by
 Dorokhov~\cite{dor},  and independently by Mello, Pereyra, and
Kumar~\cite{mpk} (for $\beta=1$) by looking at
the infinitesimal transfer matrix describing the addition
of a thin slice to the wire. The resulting evolution equation
for the eigenvalue distribution $P(\{\lambda_i\},s)$
is usually known as
 Dorokhov-Mello-Pereyra-Kumar (DMPK) equation.
The only assumptions which are
needed to obtain this equation are first
that the conductor must be weakly disordered so that the scattering in
the thin slice can be treated by using perturbation theory and second
that the flux incident in one scattering channel is, on average, equally
distributed among all outgoing channels. It is exactly this second
assumption which restricts the DMPK equation
 to the quasi 1-d  regime, where  finite time scale for transverse
diffusion can be neglected. The results of~\cite{mpk} were then
generalized to $\beta=2,4$ in Refs.~\cite{ms,mc}.
The DMPK equation is:

\begin{equation}
\frac{\partial P}{\partial s}~=
{}~D~P,
\label{DMPK}
\end{equation}
where $s$ is the length $L$ measured in units of the mean free path $l$:
$s\equiv L/l$
and $D$ can be written in terms of the $\lambda_i$'s as follows:
\begin{equation}
{}~D~=
\frac{2}{\gamma}\sum_{i=1}^{N}
\frac{\partial}{\partial\lambda_{i}}\lambda_{i}(1+\lambda_{i})
J(\lambda)\frac{\partial}{\partial\lambda_{i}}J(\lambda)^{-1},
\label{DMPK2}
\end{equation}

with $\gamma=\beta N+2-\beta$.
$\beta\in\{1,2,4\}$ is the symmetry index of the ensemble of scattering
matrices: $\beta=1$ for the orthogonal ensemble,
$\beta=2$ for the unitary ensemble
and $\beta=4$ for the symplectic one, in full analogy with the well know
Wigner-Dyson classification.
 $J(\{\lambda_{n}\})$ denotes the Jacobian from the matrix to the
eigenvalue space:
\begin{equation}
J(\{\lambda_{n}\})=\prod_{i<j}|\lambda_{j}-\lambda_{i}|^{\beta}.
\label{jacobian}
\end{equation}

There is however a completely independent, and very elegant, way to
obtain the DMPK equation. Let us assume as dynamical principle to obtain
an equation of motion for $P(\lambda)$
the simplest possible choice compatible with the
constraints described in sect.2. That is, let us assume that, as $L$
(the length of the wire) increases,
the matrix $M$ freely diffuses in the $G/H$ space going from the
perfectly conducting limit ($L=0$) (which plays the role of initial
condition for this evolution equation) to the insulating, localized
limit ($L=\infty$). In general a free diffusion in $G/H$ is described by
the Laplace-Beltrami operator of $G/H$. However, since the $\lambda$'s
are the radial coordinates of $G/H$, their behaviour as a function of
$L$ will only depend on the radial part $B$ of the Laplace-Beltrami
operator. The resulting equation is:
\begin{equation}
\frac{\partial P}{\partial s}~=~
\alpha~B~P,
\label{DMPKbis}
\end{equation}
where $\alpha$ is a (for the moment undetermined) diffusion constant
and $B$ is defined as follows:

\begin{equation}
B=[\xi(x)]^{-2}\sum_{k=1}^{n}\frac{\partial}{\partial x_k}[\xi(x)]^2
\frac{\partial}{\partial x_k}~~,
\label{defb}
\end{equation}
where we have chosen the following parametrization for the radial
 coordinates of the manifold: $\lambda_i=\sinh^2x_i$, and

\begin{equation}
\xi(x)=\prod_{i<j}|\sinh^{2}x_{j}-\sinh^{2}x_{i}|^{
\frac{\beta}{2}}
\prod_{i}|\sinh 2x_{i}|^{\frac{1}{2}}.
\label{BR2}
\end{equation}

It is now only matter of straightforward algebra to recognize that $B$
and the DMPK operator $D$ are related by:

\begin{equation}
D=\frac{1}{2\gamma}~[\xi(x)]^2~ B~ [\xi(x)]^{-2},
\label{DtoB}
\end{equation}

thus allowing, through a suitable choice of $\alpha$ and normalization
of $P(\lambda)$ to identify eq.(\ref{DMPK}) and eq.(\ref{DMPKbis}).
This identification was first recognized by  H\"{u}ffmann~\cite{huf},
and has been recently discussed in~\cite{c1} and~\cite{cm95}.

Eq.(\ref{DMPKbis}) can be considered the analogous, in the context of
the TME's (hence for symmetric spaces  of negative curvature) of the
Brownian motion approach in the case of S-matrix ensembles (which as we
shall show below are characterized, by  the same symmetric spaces, but
with positive curvature).

An important and unexpected property of the DMPK equation is that if
$\beta=2$ the various $\lambda_i$ can be decoupled.
This was recently realized by
Beenakker and Rejaei~\cite{br} who showed that the DMPK equation can be
rewritten as Schr\"odinger-like equation (in imaginary time)
for a set of $N$ interacting fermions.
The mapping was obtained by setting:
$\lambda_{n}=\sinh^{2}x_{n}$, and
 by making the following substitution
\begin{equation}
P(\{x_{n}\},s)~=~\xi(x)~\Psi(\{x_{n}\},s).
\label{BR1}
\end{equation}

In this way the DMPK equation becomes exactly equivalent to:

\begin{eqnarray}
\label{BR3}
&&-\frac{\partial\Psi}{\partial s}=({\cal H}-U)\Psi,
\label{BR3a}\\
&&{\cal H}=-\frac{1}{2\gamma}\sum_{i}\left(\frac{\partial^{2}}
{\partial x_{i}^{2}}+\frac{1}{\sinh^{2}2x_{i}}\right)\nonumber\\
&&+\frac{\beta(\beta-2)}{2\gamma}\sum_{i<j}
\frac{\sinh^{2}2x_{j}+\sinh^{2}2x_{i}}
{(\cosh 2x_{j}-\cosh 2x_{i})^{2}},\label{BR3b}
\end{eqnarray}
\eq
U=-\frac{N}{2\gamma}-N(N-1)\frac{\beta}{\gamma}-
N(N-1)(N-2)\frac{\beta^{2}}{6\gamma}.\label{BR3c}
\en
By choosing $\beta=2$, the remaining interaction terms among the $x_n$
disappear, the equation can be decoupled and can be solved
exactly~\cite{br}.

This equivalence with a Schr\"odinger equation is another feature
of the DMPK equation which has a natural explanation in the framework of
the Calogero-Sutherland models.

\section{Calogero-Sutherland Models}

 These models describe  $N$ particles on a line,
identified by their
coordinates $\{x_i\},~~i=1\cdots N$, interacting  (at
least in the simplest version of the models) with a pairwise
potential $f(x_i,x_j)$. Several realizations of this potential have
been studied in the literature  (for a comprehensive review see
ref.~\cite{op}), but in the following we shall mainly be interested in
only two realizations. The $sin$-type  CS model, in which
$f(x_i,x_j)=1/\sin^2(x_i-x_j)$ and the $sinh$-type for which
$f(x_i,x_j)=1/\sinh^2(x_i-x_j)$.

The most relevant feature of these models
is that (under particular conditions discussed below, see
eq.(\ref{CS1}) and (\ref{GM})) they have $N$
commuting integrals of motions, they are completely integrable and
their
Hamiltonian can be mapped into the radial part of a
Laplace-Beltrami operator on a
suitable symmetric space. In particular we have spaces with negative
curvature for the sinh models and of positive curvature for the sin-type
ones.

In the original
formulation of the CS model, the interaction among the particles was
simply pairwise~\cite{cs}. But it was later realized that the
complete integrability of the model had a deep group theoretical
explanation, that the simple pairwise interaction was the signature of
an underlying structure: namely the root
lattice of the Lie algebras $A_N$,
and that all the relevant properties (complete integrability, mapping
to a Laplace-Beltrami operator of a suitable symmetric space) still
hold for potentials constructed by means of any root lattice
canonically associated to a simple Lie algebra~\cite{op}.
Let us see  more precisely how this construction works.

Let us call ${\cal V}$ the $N$ dimensional space defined by the
coordinates $\{x_i\}$ and $x=(x_1,\cdots x_N)$ a vector in ${\cal V}$.
Let $R=\{\alpha\}$ be a root system in ${\cal V}$, and $R_+$  the
subsystem of positive roots of $R$. Let us denote with $x_\alpha$
the scalar product $(\alpha,x)$.
Then the general form of the CS Hamiltonian is
\begin{equation}
H=-\frac{1}{2}\sum_{i=1}^{N}\frac{\partial^2}{\partial^2x_i}+
\sum_{\alpha\in R_+}\frac{g^2_\alpha}{\sinh^2(x_\alpha)}
\label{CS1}
\end{equation}
 where the couplings $g^2_\alpha$ are the same for equivalent roots,
namely for those roots which are connected with each other by
transformations of the Coxeter group $W$ of the root system.
To clarify this rather abstract definition let
us see two examples, obtained using the root lattices $A_N$ and
$C_N$ (in the following $\{e_i,\cdots e_N\}$ denote a
canonical basis in the space {\bf R}$^n$).
\begin{description}
\item{$A_N$:~~} This root system is obtained by taking a hyperplane in
${\bf R}^{N+1}$ for which $x_1+x_2+\cdots x_{N+1}=1$. Then the root
system $R$ is given by:
  $R=\{e_i-e_j,~i\neq j\}$. In this
case W is the permutation group of the set $\{e_i\}$. The corresponding
Hamiltonian is:

\begin{equation}
H=-\frac{1}{2}\sum_{i=1}^{N}\frac{\partial^2}{\partial^2x_i}+
\sum_{i<j}\frac{g^2}{\sinh^2(x_i-x_j)}
\label{CS2}
\end{equation}

This is the model originally considered in~\cite{cs}

\item{$C_N$:~~} This root system is $R=\{\pm 2e_i,~\pm e_i \pm e_j,
{}~i\neq j\}$, in this
case W is the product of the permutation group and the group of
transformations which change the sign of the vectors $\{e_i\}$.
The corresponding Hamiltonian is:

\begin{eqnarray}
H=-\frac{1}{2}\sum_{i=1}^{N}\frac{\partial^2}{\partial^2x_i}+
\sum_i\frac{g_2^2}{\sinh^2(2x_i)}&&\nonumber\\
+\sum_{i<j}\left(\frac{g_1^2}{\sinh^2(x_i-x_j)}
+\frac{g_1^2}{\sinh^2(x_i+x_j)}\right)&&
\label{CS3}
\end{eqnarray}
This is the model which we shall study in the following.

\end{description}

By using  simple identities among hyperbolic functions
eq.(\ref{CS3}) can be rewritten as follows:
\begin{eqnarray}
H&=&-\frac{1}{2}\sum_{i=1}^{N}\frac{\partial^2}{\partial^2x_i}
+\sum_i\frac{g_2^2}{\sinh^2(2x_i)}+c\nonumber\\
&+&2g_1^2\sum_{i<j}
\frac{\sinh^2(2x_i)+\sinh^2(2x_j)}
{\left(\cosh(2x_i)-\cosh(2x_j)\right)^2}
\label{CS4}
\end{eqnarray}
with $c$ an irrelevant constant.
By setting
\eq
g_2^2=-1/2~,~~~~~  g_1^2=\frac{\beta(\beta-2)}{4}
\label{4.b}
\en
 we see that
eq.(\ref{CS3}) coincides (apart from the overall factor $1/\gamma$)
with ${\cal H}$ in eq.(\ref{BR3b}).

As we mentioned above, the relevant feature of the CS hamiltonian
(\ref{CS3}) is that it can be mapped into the radial part $B$ of
a Laplace-Beltrami operator
(see for instance Appendix D of ref.~\cite{op}) of a suitable symmetric
space
\begin{equation}
H=\xi(x)\left[\frac{1}{2}(B+\rho^2)\right]\xi(x)^{-1}
\label{4.a}
\end{equation}
with $B$ defined by eq.(\ref{defb})
and $\rho$ a constant term which we shall neglect in the following.
The particular symmetric space is uniquely fixed by the root lattice
underlying the CS Hamiltonian and by the coupling constant $g_\alpha$
In fact it is well known that all
the irreducible
symmetric spaces of classical type can be classified
with essentially the same techniques used for the Lie algebras.
They fall into 11 classes labelled by the type of root system and by the
multiplicities of the various roots~\cite{helg}. Some of these spaces
(those relevant for our discussion) are listed in tab.2 and 3
 with their root multiplicities.

\begin{table}[htb]
\small
\begin{center}
\begin{tabular}{|c|c|c|}
\hline
 $G$ & $H$  &  $\beta$ \\
\hline
$SL(N,{\bf R})$ & $SO(N)$   & 1  \\
$SL(N,{\bf C})$ & $SU(N)$   & 2  \\
$SU^*(N)$ & $USp(2N)$   & 4  \\
\hline
\end{tabular}
\end{center}
\caption[h]
        {\label{an}
        Irreducible symmetric spaces of type $A_N$.
        In the first two columns the group $G$ and the maximal
        subgroup $H$ which define the symmetric space.
         All these spaces are
        labelled by the Dynkin diagram $A_{N-1}$. In the last column the
        multiplicity $\beta$ of the roots}
\end{table}

\begin{table}[htb]
\small
\begin{center}
\begin{tabular}{|c|c|c|c|}
\hline
 $G$ & $H$  & $\beta$ & $\eta$ \\
\hline
$Sp(2N,\mbox{\bf R})$ & $U(N)$  & 1 & 1  \\
$SU(N,N)$ & $SU(N)\otimes SU(N)\otimes U(1)$   & 2 & 1  \\
$SO^*(4N)$ & $U(2N)$  & 4 & 1  \\
$USp(2N,2N)$ & $USp(2N)\otimes USp(2N)$   & 4 & 3  \\
$Sp(2N,\mbox{\bf C})$ & $USp(2N)$  & 2 & 2  \\
\hline
\end{tabular}
\end{center}
\caption
        {\label{cn2}
        Irreducible symmetric spaces of type $C_N$.
        In the first two columns the group $G$ and the maximal
        subgroup $H$ which define the symmetric space.
         All these spaces are
        labelled by the Dynkin diagram $C_{N}$. In the third column the
        multiplicity $\beta$ of the ordinary roots of $C_N$. In the last
        column the multiplicity $\eta$ of the long root.}
\end{table}

These multiplicities are related to the coupling constants by~\cite{op}
\begin{equation}
g_\alpha^2=\frac{m_\alpha(m_\alpha-2)}{8}|\alpha|^2
\label{GM}
\end{equation}
where $|\alpha|$ is the length of the root $\alpha$ and $m_\alpha$ its
multiplicity.
Only for these special values of $g^2_\alpha$ the mapping (\ref{4.a}) is
possible.
For the three ensembles in which we are interested we have
$m_\alpha=\beta$ for the short roots (those of the
type $\{\pm e_i \pm e_j\}$) and $m_\alpha=1$ for the long roots
(those of the type $\{2 e_i\}$), which if inserted in eq.(\ref{GM})
exactly give the values of eq.(\ref{4.b}).
In this way we see that the index $\beta$ has a deep group theoretical
meaning, since it denotes the  multiplicity of the (ordinary) roots of
the symmetric space in which the transfer matrix diffuses as $L$
increases. The mapping described by eq.(\ref{4.a}) is exactly the one
found in ref.~\cite{br}.

All the symmetric spaces listed in tab.2 and 3 are spaces of negative
curvature. For each one of them there is a counterpart of positive
curvature, with all other properties (in particular the root lattice
structure) unchanged. For instance we have:
\eq
\frac{SL(N,{\bf R})}{SO(N)} \to
\frac{SU(N)}{SO(N)}
\en
In the context of CS models one moves from negative to positive
curvature symmetric spaces by changing the sinh-type interaction into the
sin-type one. In the framework of RMT for disordered wires one has the
same change moving from the TME to the S-matrix ensembles. In fact, as
we mentioned above, the flux conservation constraint implies that the $S$
matrix belongs  to a compact group. At this point,
depending on the problem in which
one is intersted, and consequently, on the  parametrization which one
choses for the (possibly generalized) eigenvalues one can find WDE's,
which correspond to sin CS models of $A_N$ type, or the S-matrix
ensembles
described in~\cite{fp} which correspond to sin CS model of $C_N$ type.

\section{solution of the DMPK equation}

The most important application of the above described equivalence
between TME and CS models is that, by using some recent results on the
CS models  one can obtain several important properties of the DMPK
equation: solve it exactly in the case $\beta=2$ and find approximate
asymptotic solutions for $\beta=1,4$. This result was obtained
 in~\cite{c1} and we shall describe here the main steps of that solution.
According to eq.(\ref{DtoB}) if $\Phi_k(x)$, $x=\{x_1,\cdots,x_N\}$,
$k=\{k_1,\cdots,k_N\}$ is an eigenfunction of
$B$ with eigenvalue $k^2$, then $\xi(x)^2\Phi_k(x)$ will be an
eigenfunction of the DMPK operator with eigenvalue $k^2/(2\gamma)$.
These eigenfunctions  of the $B$  operator  are known in the literature
as ``zonal spherical functions''. In the following we shall use
three important properties of these functions (see~\cite{hc}).

\vskip 0.2cm
\noindent
{\bf 1}] By means of the zonal spherical functions one can define the
analog of the Fourier transform on symmetric spaces:
\begin{equation}
f(x)=\int \bar f(k) \Phi_k(x) \frac{dk}{|c(k)|^2}
\label{ss1}
\end{equation}
(where we have neglected an irrelevant multiplicative constant)
and in the three cases which are of interest for us:
\begin{equation}
|c(k)|^2=|\Delta(k)|^2
\prod_{j}\left\vert\frac{\Gamma\left(
i\frac{k_j}{2}\right)}{\Gamma\left(\frac{1}{2}+
i\frac{k_j}{2}\right)}\right\vert^2
\label{ss2}
\end{equation}
with
\begin{equation}
|\Delta(k)|^2=\prod_{m<j}\left \vert\frac{\Gamma\left(
i\frac{k_m-k_j}{2}\right)\Gamma\left(
i\frac{k_m+k_j}{2}\right)}{\Gamma\left(\frac{\beta}{2}+
i\frac{k_m-k_j}{2}\right)\Gamma\left(\frac{\beta}{2}+
i\frac{k_m+k_j}{2}\right)}\right\vert^2
\label{ss2b}
\end{equation}
where $\Gamma$ denotes the Euler gamma function.

\vskip 0.2cm
\noindent
{\bf 2}] for large values of $x$,
$\Phi_k(x)$ has the following asymptotic behaviour:
\begin{equation}
\Phi_k(x)\sim \frac{1}{\xi(x)} \left(\sum_{r\in W} c(r k) e^{i(r k,x)}
\right)~~,
\label{ss3}
\end{equation}
where $r k$ is the vector obtained acting with $r\in W$ on $k$.
The important feature of eq.(\ref{ss3}) is that it is valid for {\it all
values of k}.

\vskip 0.2cm
\noindent
{\bf 3}]
in the case $\beta=2$ the explicit form of $ \Phi_k(x)$
is known~\cite{op,bk}:
\begin{equation}
\Phi_k(x)=\frac{det\left[ Q_m^j\right]}
{\prod_{i<j}[(k_i^2-k_j^2)(\sinh^2 x_i
-\sinh^2 x_j)]}
\label{ss4}
\end{equation}
where the matrix elements of $Q$ are:
\begin{equation}
Q_m^j={\rm F}\left(\frac{1}{2}(1+ik_m),
\frac{1}{2}(1-ik_m),1;-\sinh^2x_j\right)
\label{ss4b}
\end{equation}
and $F(a,b,c;z)$ is the hypergeometric function.

\vskip 0.2 cm
Eq.s(\ref{DtoB},\ref{ss1}-\ref{ss2b}) allow to write the $s$-evolution of
$P(\{x_n\},s)$  from  given initial conditions (described by the
function $\bar f_{0}(k)$) as follows:
\begin{equation}
P(\{x_n\},s)=[\xi(x)]^2 \int \bar f_0(k)e^{-\frac{k^2}{2\gamma}s}
 \Phi_k(x) \frac{dk}{|c(k)|^2}.
\label{ss5}
\end{equation}
 By inserting the explicit expression of
$|c(k)|^2$ and by using the identity:
\begin{equation}
\left\vert\frac{\Gamma\left(\frac{1}{2}+
i\frac{k}{2}\right)}
{\Gamma\left(
i\frac{k}{2}\right)}\right\vert^2=\frac{k}{2}\tanh\frac{\pi k}{2}
\end{equation}
we end up with the following general expression for $P(\{x_n\},s)$
with ballistic initial conditions (which, due to the normalization
of $\Phi_k(x)$, simply amount to choosing $ \bar f_0(k)=const$):

\eqa
&&P(\{x_n\},s)=[\xi(x)]^2 \int dk
 e^{-\frac{k^2}{2\gamma}s}
\frac{ \Phi_k(x)}{|\Delta(k)|^2} \nonumber \\
&& \prod_{j}k_j\tanh(\frac{\pi k_j}{2})
\label{ss6}
\ena
This expression is rather abstract, but it can be made more explicit by
using the properties [2] and [3] listed above.
In the $\beta=2$ case we can insert the explicit expression
for $\Phi_k(x)$, given in eq.s(\ref{ss4},\ref{ss4b}),
into eq.(\ref{ss6}). By  using the identity
\begin{equation}
{\rm P}_{\nu}(z)=F(-\nu,\nu+1,1;(1-z)/2)
\end{equation}
 we exactly
obtain (as expected) the solution,  found by Beenakker and
Rejaei in the same case~\cite{br}.

In the other two cases $\beta=1,4$,
if $x$ is large (and in our framework this means $x^2> (2s)/
\gamma$) we may insert  the asymptotic
expansion (\ref{ss3}) into eq.(\ref{ss6}).
 The resulting behaviour of $P(\{x_{n}\},s)$  will depend
on the chosen (metallic or insulating) regime for $k$.
Let us look at the two cases  separately.

\vskip 0.2cm

{\it Insulating regime $(k\ll 1)$.}

\noindent
In the $k \to 0$ limit the $\Gamma$ functions in eq.(\ref{ss3})
can be approximated according to:
\begin{equation}
\frac{\Gamma(\frac{\beta}{2}+iy)}{\Gamma(iy)}\sim_{y\to 0}
iy~~,~~\beta\in\{1,2,4\}~.
\end{equation}
Then, by
rewriting both the
product $\prod_{i<j}(k^2_i-k^2_j)$, and the sum over the exponentials in
(\ref{ss3}) as  determinants,
the  integration over $k$ becomes straightforward and gives:

\begin{eqnarray}
P(\{x_{n}\},s)=\prod_{i<j}\left\vert\sinh^{2}x_{j}-\sinh^{2
}x_{i}\right\vert^{\frac{\beta}{2}}\left[
(x_{j}^{2}-x_{i}^{2})\right]&&\nonumber\\
\times \prod_{i}\left[\exp(-x_{i}^{2}\gamma/(2s))x_{i}(\sinh
2x_{i})^{1/2}\right].&&
\label{f1}
\end{eqnarray}

 Ordering the
$x_{n}$'s from small to large and using the fact that in this regime
 $1\ll x_{1}\ll x_{2}\ll\cdots\ll x_{N}$ we can
approximate the eigenvalue distribution as follows:
\begin{equation}
P(\{x_{n}\},s)=\prod_{i=1}^{N}
\exp\left[-(\gamma/(2s))(x_{i}-\bar{x}_{i})^{2}\right].
\label{IR2}
\end{equation}
where $\bar{x}_{n}=\frac{s}{\gamma}(1+\beta(n-1))$,
in agreement with the result obtained by Pichard,\cite{pic} by directly
solving the DMPK equation in this regime.

\vskip 0.2cm

{\it Metallic regime $(k\gg 1)$.}

\noindent
In this case one must use
the asymptotic expansion:
\begin{equation}
\frac{\Gamma(\frac{\beta}{2}+iy)}
{\Gamma(iy)}\sim_{y\to\infty} |y|^{\frac{\beta}{2}}~e^
{\frac{i\pi\beta}{4}}~,~~\beta\in\{1,2,4\}~.
\end{equation}
The integration over $k$ is less simple in this case and,
(to be consistent with the regime of validity of eq.(\ref{ss3}))
in the resulting expression
only the highest powers of $(x\sqrt{\frac{\gamma}{2s}})$
must be taken into account.
We find:
\begin{eqnarray}
P(\{x_{n}\},s)=\prod_{i<j}\left\vert\sinh^{2}x_{j}-
\sinh^{2}x_{i}\right\vert^{\frac{\beta}{2}}\left\vert
x_{j}^{2}-x_{i}^{2}\right\vert^{\frac{\beta}{2}}&&\nonumber\\
\times\prod_{i}\left[\exp(-x_{i}^{2}\gamma/(2s))(x_{i}\sinh
2x_{i})^{1/2}\right].&&
\label{f2}
\end{eqnarray}
In agreement with the exact result of ~\cite{br} for $\beta=2$ and with
the $\beta$ dependence found by Chalker
and Mac\^{e}do~\cite{cm2} through a direct integration of the DMPK
equation.

According to  ref.~\cite{op} and ref.~\cite{hc}, the regime of validity
of eq.s(\ref{f1}) and (\ref{f2})  is  $x^2> (2s)/\gamma$, which gives in
in the large $N$ limit $x^2>(2s)/(\beta N)$.
Notice however
that eq.(\ref{ss3}) is only the first term of a series
which converges absolutely to  $\Phi_k(x)$ for all values of $k$ (see
sect. 8 of ref.~\cite{op} and ref.~\cite{hc}).
The coefficients of
this series can be constructed recursively, thus allowing
to study the behaviour of eq.(\ref{ss6}) even for values of $x$ smaller
than the above mentioned threshold.

\section{Conclusions and Perspectives}

We have shown that there is a deep connection between CS models and the
RMT approach to disordered wires. In particular we have shown that
sinh CS models of $C_N$ type correspond to TME's, that
sin CS models of $C_N$ type correspond to the S-matrix ensembles in the
parametrization of ref~\cite{fp} and that ordinary WDE's correspond to
sin CS models of $A_N$ type. Several properties of the TME's can be
obtained by exploiting this connection. In particular exact or
approximate solution of the DMPK equation can be obtained.

Another interesting application of this correspondence can be found in
the  insulating regime.  Looking at tab.2 and tab.3 we see that
the most relevant feature of the $C_N$ symmetric spaces with respect of
the $A_N$ ones is that they are described by two critical indices
instead of one. This is due to the fact that the $C_N$ Dynkin diagrams
have two types of roots, each with its particular multiplicity. We shall
call this second index $\eta$ in the following. This peculiar feature of
these ensembles is usually ignored  because in the weakly
localized regime it is only the value of $\beta$ which matters, and also
because the three cases which have been
 studied up to now (the first three lines of tab.1) have the same
value of the second index $\eta=1$, thus leading to the same DMPK
equation (with no explicit $\eta$ dependence).

However in the insulating regime also the index $\eta$ becomes important
and can be directly measured by looking for instance at the
ratio~\cite{c2}

\eq
\frac{var~(\log{g})}{<\log{g}>}=\frac{2}{\eta}.
\en

In fact, while in the weakly localized
regime all the relevant physical properties are completely determined
by the level statistics  (namely the value of $\beta$),
 when extending the RMT approach to the insulating regime, due to the
fact that in this case {\it the conductance is  dominated by the lowest
eigenvalue} all the details of the chosen RMT model
(number of degrees of freedom, possibly the presence of new critical
indices) become important~\cite{c2}. This observation could open a new
interesting field of application of the Calogero-Sutherland techniques
in the physics of disordered wires

\vskip 0.3 cm
{\bf Acknowledgements}
I want to warmly thank G. Zemba and F. Gliozzi for their constant help
and encouragement.

\end{document}